\documentclass[9pt,twocolumn,twoside]{opticajnl}
\journal{opticajournal} % use for journal or Optica Open submissions

\setboolean{shortarticle}{true}

\usepackage{lineno}
\usepackage{upgreek}
\usepackage{xcolor}

\usepackage{wrapfig}
\usepackage{graphicx}
\usepackage{lipsum}
\usepackage{subcaption}
\usepackage{siunitx}
\usepackage{ulem}
\usepackage{siunitx}
\usepackage{xcolor}

\begin{document}

\title{Directly measured squeeze factors over GHz bandwidth from monolithic ppKTP resonators}

\author[1,*]{Benedict Tohermes}
\author[1,*]{Sophie Verclas}
\author[1]{Roman Schnabel}

\affil[1]{Institut f\"ur Quantenphysik, Universit\"at Hamburg, Luruper Chaussee 149, 22761 Hamburg, Germany}
\affil[*]{Shared first authorship}

\begin{abstract}
Squeezed vacuum states of light with bandwidths in the gigahertz range are required for ultrafast quantum sensors, for high-bandwidth QKD and for optical quantum computers. Here we present squeeze factors of monolithic periodically poled KTP (ppKTP) resonators measured with two laboratory-built balanced homodyne detectors with gigahertz bandwidth. We realise two complete systems without selection of optical or electronic hardware components to test the reproducibility without rejects.
As expected, the systems show clear spectral differences. However, both achieve directly measured squeeze factors in the order of 3 dB over a GHz bandwidth, which is achieved here for the first time. Our direct measurement of quantum correlation is suitable for increasing the key rate of one-sided, device-independent QKD.
\end{abstract}

\maketitle

\section{Introduction}
\label{sec:introduction}

The first measurements of squeezed light from resonator-enhanced parametric down-conversion were performed in 1987 \cite{Wu1987}. The squeeze factors were subsequently improved \cite{Polzik1992, Takeno2007, Vahlbruch2008, Schnabel2017} until up to 15\,dB were measured in 2016 at 1064\,nm \cite{Vahlbruch2016}. Squeezed light has applications in a wide field of applications, including quantum sensing \cite{LSC2011,Taylor2013,Ganapathy2023,Yu2023, Zander2023, Korobko2023, Wolfgramm2010}, optomechanics \cite{Hoff2013,Kleybolte2020} and quantum key distribution (QKD) \cite{Gehring2015} and optical quantum computing \cite{Larsen2019}.  
An efficient way to realise a QKD link is to distribute quantum states of light at 1550\,nm over a fibre network.
It was shown that two-mode squeezed Gaussian Einstein-Podolsky-Rosen (EPR) entangled states \cite{Furusawa1998,Eberle2013,Zander2021} enable one-sided device-independence \cite{Gehring2015,Furrer2012}.
Squeezed states are usually measured in the same way as coherent states (unconditioned continuously and with unity photoelectrical gain), but show a higher decoherence under loss than these. 
However, when using low-loss conventional optical fibres and carrier light at a wavelength of 1550\,nm, a positive key rate over a distance of several kilometres is easily achievable \cite{Gehring2015}. The key rate scales linearly with the bandwidth of the distributed entangled (squeezed) states. This motivates not so much a high squeeze factor as a high bandwidth of the squeezed sideband spectrum. If possible, it should be as high as the bandwidth of available photoelectric detectors.

Today, `squeeze lasers' \cite{Schnabel2017,Schnabel2022} providing free-space beams with tens of MHz squeeze bandwidth in one longitudinal mode are commercially available \cite{NoisyLabs}. The bandwidth is limited by the linewidth of the squeezing resonator. Squeezing resonators with several hundreds of megahertz and even gigahertz sideband frequency were first realized in 2012 \cite{AstS2012} and improved in 2013 \cite{AstS2013}. They observed GHz bandwidth with squeeze factors of up to 4.8\,dB at 5\,MHz, 2\,dB at 100\,MHz and 1\,dB at 1.2\,GHz in a freely propagating laser beam in well-defined TEM00 mode. In more recent work, up to 6.3\,dB squeezed vacuum states were measured with THz bandwidth \cite{Kashiwazaki2020,Kashiwazaki2021} from a waveguide system.To enable measurement with such a high bandwidth, the squeezed quadrature was optically parametrically amplified into the semi-classical range. The need for low-noise electronic amplification of this bandwidth could thus be avoided. It is currently not clear, however, whether the current security proof for one-sided device-independent QKD \cite{Furrer2012} can be extended to include this measurement technique. 

Here we improve the work in \cite{AstS2012, AstS2013}. First, we reduce the optical loss experienced by the squeezed states due to imperfect measurement, namely by significantly improving the mode matching of the squeezed beam with the local oscillator beams of the balanced homodyne detectors (BHDs). On the other hand, we reduce the electronic dark noise of the transimpedance amplified BHDs over a bandwidth of up to 1.5 GHz below the optical quantum noise by a more compact design. We directly measure quantum noise reductions of up to 6.5\,dB at 30\,MHz and up to 2.9\,dB at 1\,GHz. All values given are not dark noise corrected.

\section{Experimental Setup}
\label{experimental-setup}

\begin{figure}
	\centering
	\includegraphics[width=\columnwidth]{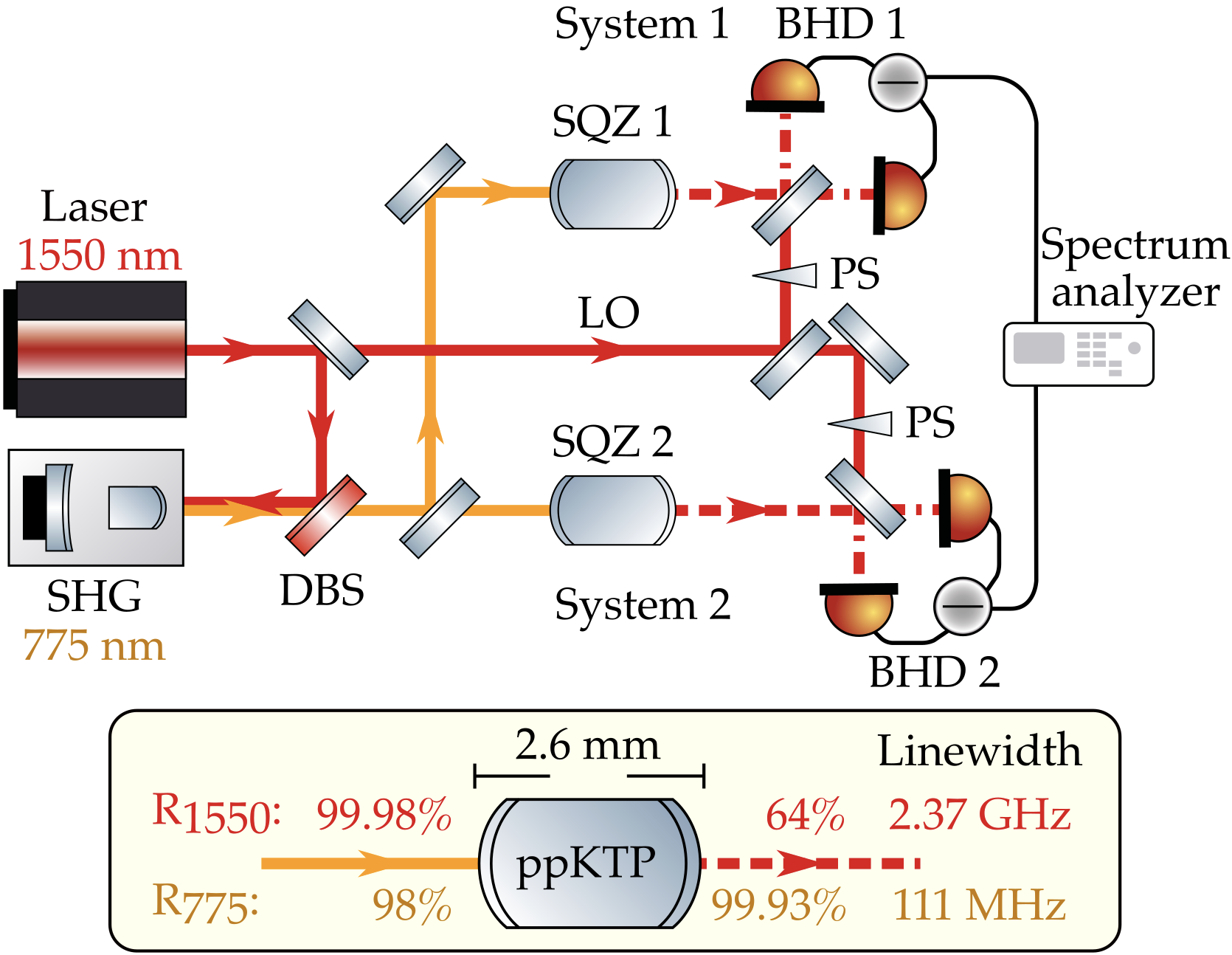}
	\caption{
Schematic drawing of our doubled squeeze laser assembly. The primary laser provided up to 2\,W output power at 1550\,nm wavelength. Its light was split up and a small part was used as local oscillator (LO) beams for the balanced homodyne detectors (BHDs). The remaining light power was sent to the second-harmonic generation resonator (SHG) and converted to 0.9\,W of 775\,nm pump light for the squeezing resonators. Squeeze laser SQZ\,1 (2) was pumped with approximately 0.6 (0.3)\,W. The noise powers of the photo-electric voltages of the BHDs were analysed with a spectrum analyzer. PS: Phase shifter, DBS: Dichroic beam splitter.}
	\label{fig:1}
\end{figure}

Figure \ref{fig:1} shows a schematic of our setup. It is similar to the setup described in \cite{AstS2013}. Especially the 2.6\,mm long crystals including coatings for forming monolithic resonators
were from the same manufacturing batch. The initial laser was a seed laser with an amplifier from NKT. It produced up to 2\,W light power at 1550\,nm. Most of the light power was coupled into a second harmonic generator (SHG), where it got converted to 775\,nm pump light for the squeeze lasers. A small fraction of its light power worked as the local oscillator for balanced homodyne detection (BHD). A detailed description can be found in \cite{AstS2012,AstS2013}. 

In contrast to \cite{AstS2012,AstS2013}, we placed the crystals on three temperature-controlled pads instead of one. All three were PID-controlled to slightly different set-point temperatures by Peltier elements. We had planned to use the middle temperature where the beam waist is for optimizing the phase matching and the outer temperatures to stabilize the crystal length to double resonance at 1550\,nm and 775\,nm.
While this scheme was a successful approach in previous experiments with longer crystals \cite{Hagemann2024}, we could not realize a significant temperature gradient inside the short crystals used here. While squeeze laser 1 showed rather high parametric gain when applying a uniform temperature profile, squeeze laser 2 did not show double resonance and good quasi-phase matching with any accessible temperature profile.
However, we achieved a significantly improved modematching of the two squeezed beams to the local oscillators of the two balanced homodyne detectors (BHD). While the mode-matching efficiency obtained in \cite{AstS2012} was only 0.81, we achieved values of approximately 0.97 by using control beams that were transmitted through the (strongly under-coupled) monolithic crystal resonators and not reflected by them from the opposite side. The optical loss associated with mode mismatch at the detector of the squeezed beams was thus reduced from 19\% to 3\%.
Also in contrast to \cite{AstS2012,AstS2013}, we used two newly self-designed and self-assembled BHDs, again with high quantum efficiency photo diodes with diameters of 100\,\textmu m in combination with high speed electronic components. Compared to our previous ones used in \cite{AstS2012} and \cite{AstS2013}, the new ones are characterized by better impedance matching, smaller dimensions and a four-layer design of the printed circuit boards.

\section{Results}\label{results}
\begin{figure}
	\centering
	\begin{subfigure}{\columnwidth}
		\includegraphics[width=0.95\columnwidth]{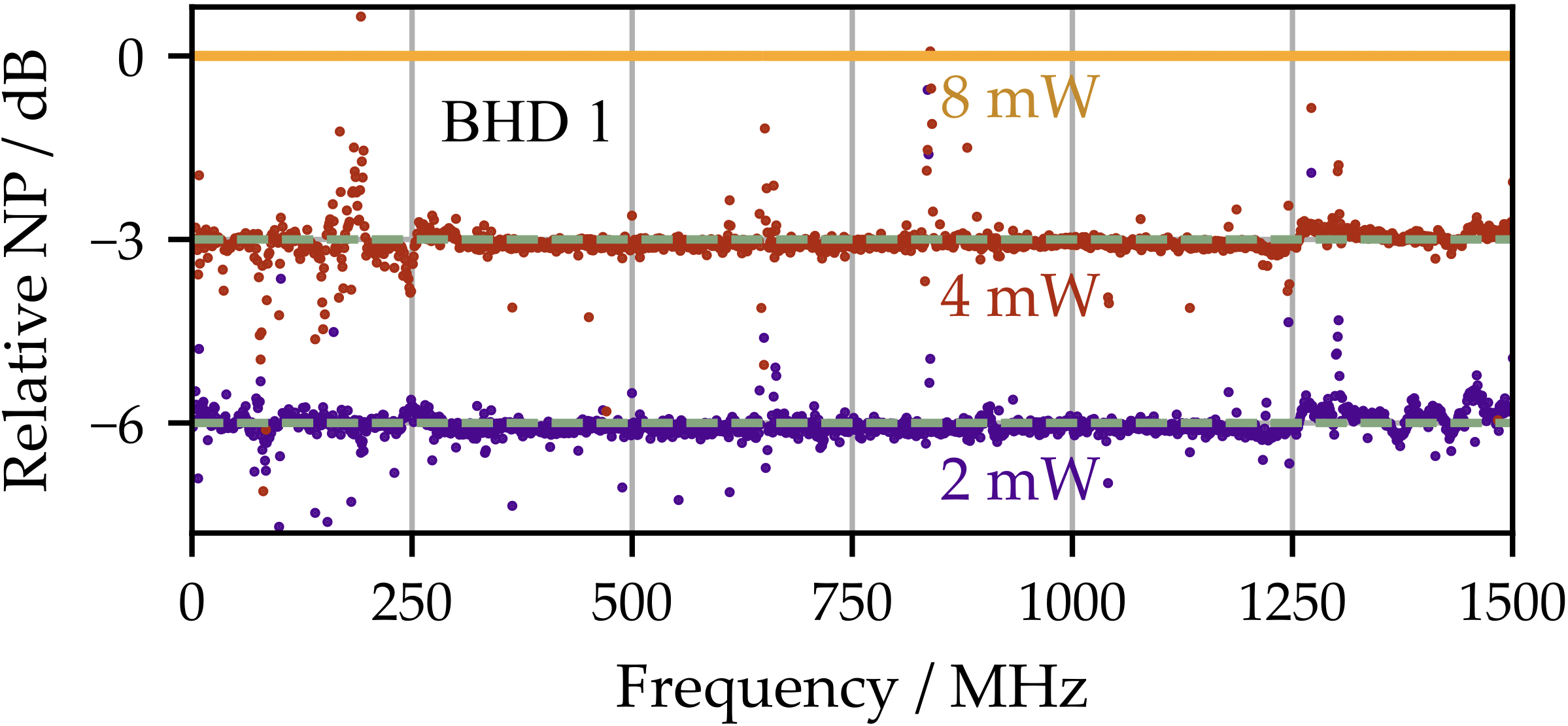}
	\end{subfigure}
	\begin{subfigure}{\columnwidth}
		\includegraphics[width=0.95\columnwidth]{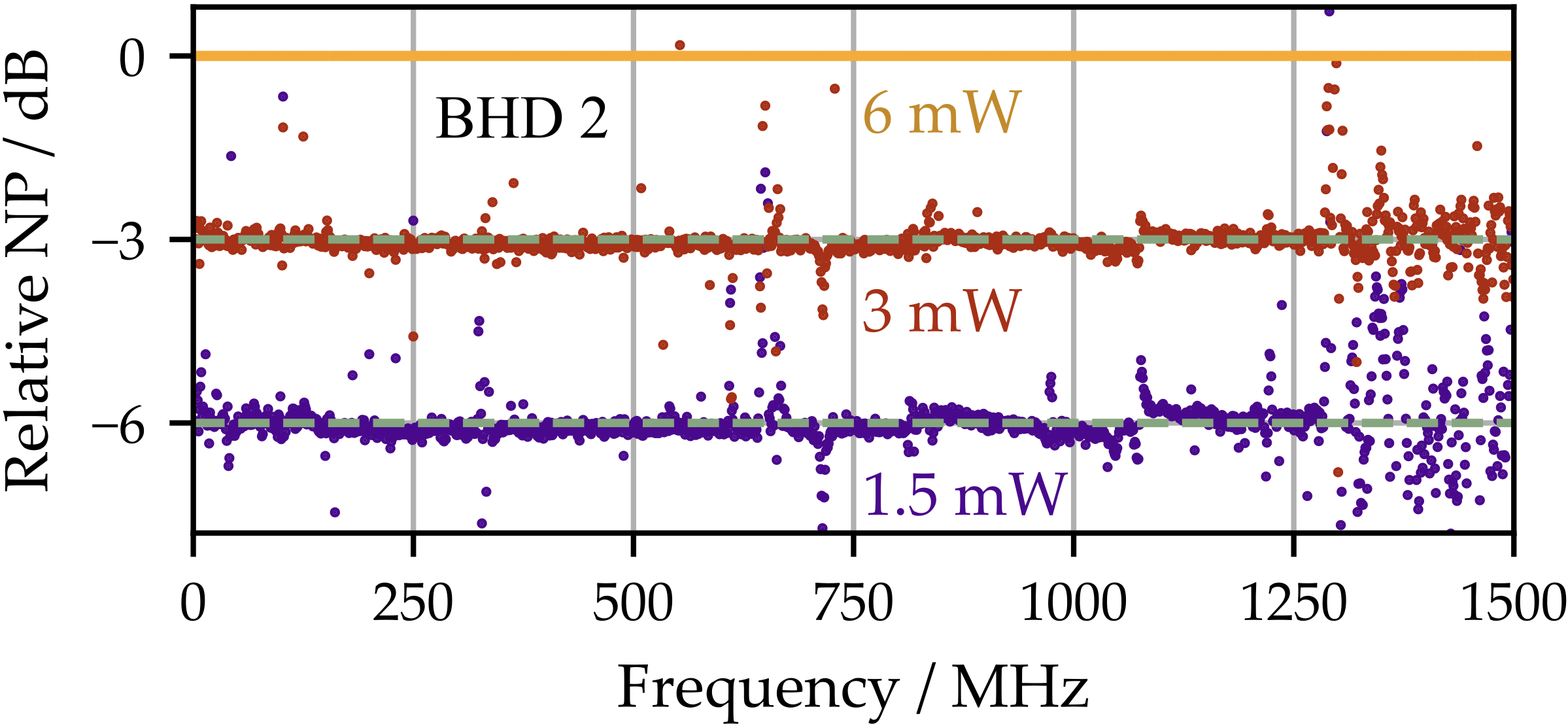}
	\end{subfigure} \vspace{-5mm}
	\caption{Relative spectral shot-noise powers (NPs) measured with BHD\,1 (top) and BHD\,2 (bottom)  for three local oscillator powers when the (squeezed) signal inputs were blocked. Both detectors showed a 3\,dB increase in noise power when doubling the local oscillator power up to 8\,mW and 6\,mW respectively, as required for the desired linear quantum noise limited operation. Narrow-band deviations were caused by electronic pickup.}
	\label{fig:2}
\end{figure}

Both systems of squeeze laser and transimpedance amplified balanced homodyne detector provide direct measurements of broad-band squeezed quantum noise with about 1.5\,GHz bandwidth. The two systems are built according to the same design, however, show different properties due to unspecified differences in crystal cuts and unspecified differences in the low-noise high-speed electronic components used. 

Figure\;\ref{fig:2} first shows the successful linearity tests of our BHDs over a factor of four of the local oscillator (LO) power. All traces shown were measured with an empty (blocked) signal input of the BHDs. In this case, the noise power of a BHD should be proportional to the variance of the ground state uncertainty of the electric field, linearly amplified by the power of the superimposed local oscillator field (3\,dB per factor two of the LO power). If the measured noise power were limited solely by the technical noise of the local oscillator beam, the measured noise power would scale quadratically with the power of the local oscillator beam (6\,dB per factor of two in the LO power). For the evaluation, the dark noise of the detectors is subtracted from the measurement data (unlike measurement data that is intended to quantify a squeeze factor). Increasing the light output above the range shown resulted in a too low increase in noise power, which indicated saturation. Reducing the light output below the range shown resulted in more noisy data due to the subtracted dark noise. Based on this characterisation, the two BHDs were operated with six and eight mW local oscillator power respectively for the following measurements.

\begin{figure}[h]
	\centering
	\begin{subfigure}{1\columnwidth} 
		{\includegraphics[width=\columnwidth]{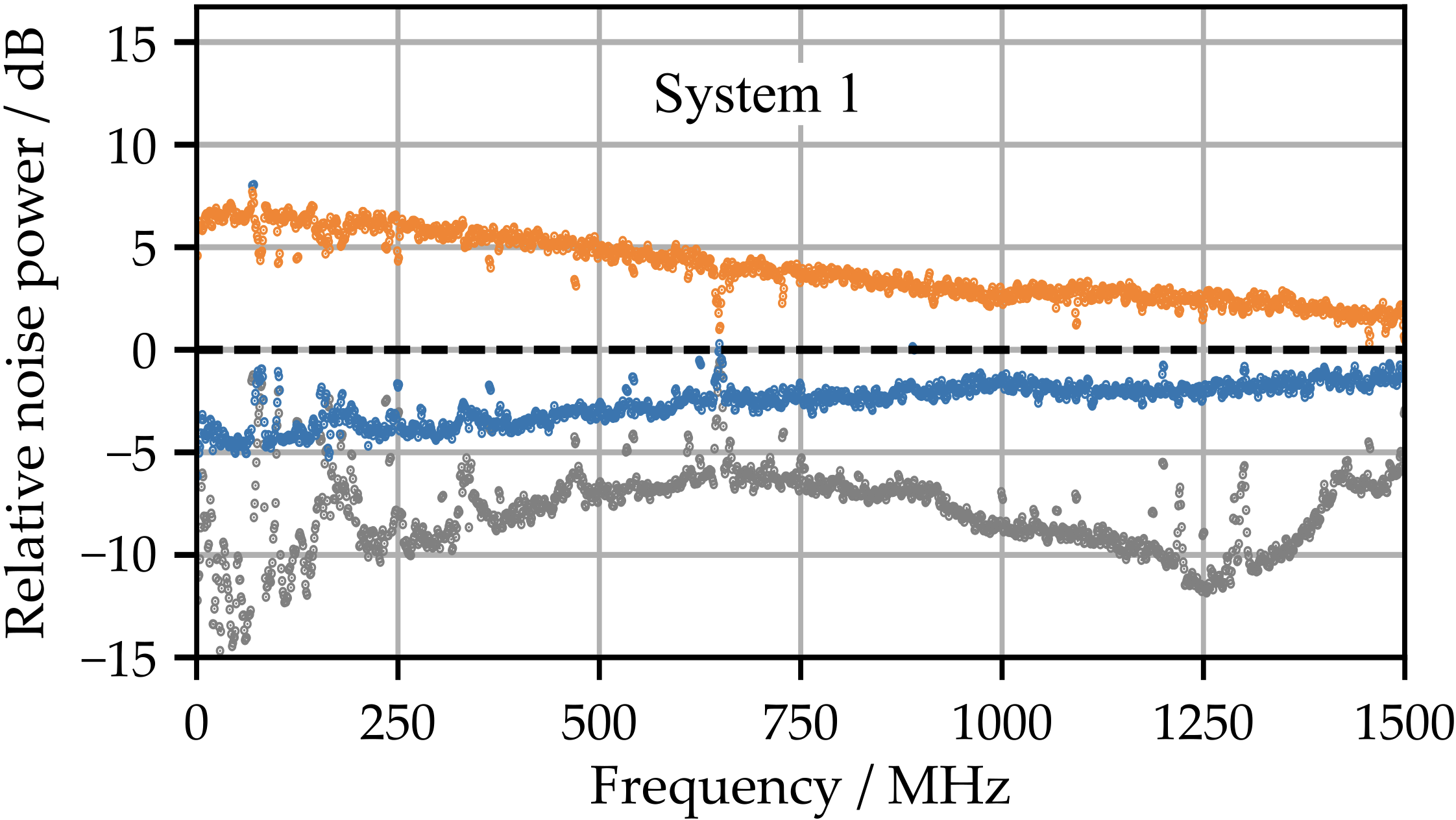}}
	\end{subfigure}
	\\[3mm]
	\begin{subfigure}{1\columnwidth} 
		{\includegraphics[width=\columnwidth]{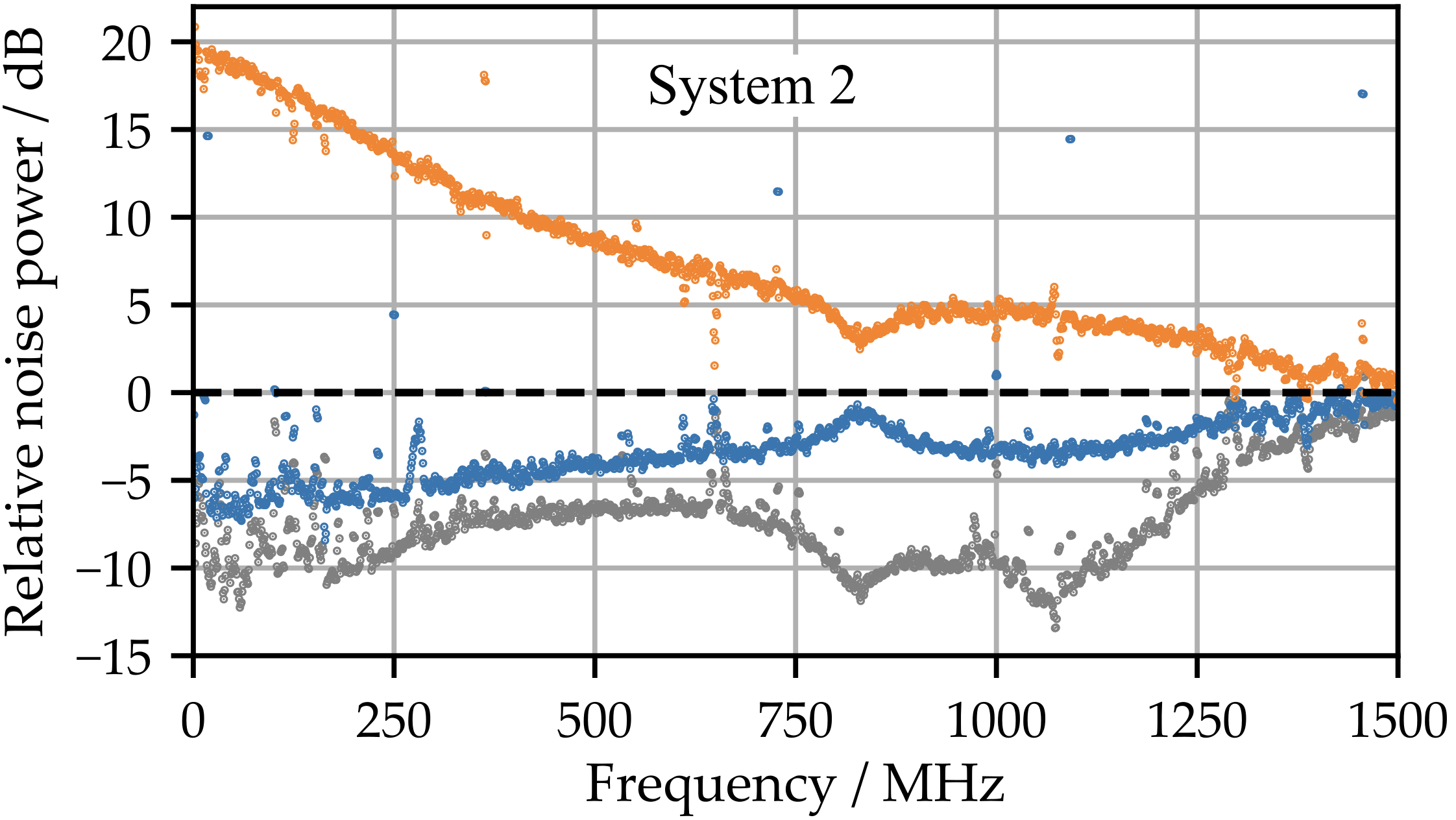}}
	\end{subfigure}  \vspace{-5mm}
	\caption{Normalized spectral quantum noise powers measured on the beams from SQZ\,1 and 2. Anti-squeezed noise power (orange, top), vacuum noise reference (dashed), squeezed noise power (blue), and BHD dark noise power (grey, bottom). Both squeeze lasers generated squeezed noise in the sideband up to 1.5\, GHz. The quantum noise squeezing measured on SQZ\,1 with BHD\,1 was mainly limited by the too low effective optical non-linearity of SQZ\,1. The quantum noise squeezing measured on SQZ\,2 with BHD\,2 was mainly limited by dark noise, an electronic resonance at approximately 0.83\,GHz and the imperfect detection efficiency of BHD\,2.}
	\label{fig:3}
	\vspace{2mm}
\end{figure}

Figure\;\ref{fig:3} presents the characterization of the quantum noise squeezing from SQZ\,1 and 2 measured with BHD\,1 and 2, respectively. The data from system\,2 show a measured noise reduction of slightly less than 5\,dB at the lower ten MHz sideband frequencies. The anti-squeezed noise of the same states was measured with the local oscillator phase-shifted by a quarter wavelength. The noise power here was up to 7\,dB higher than that of the vacuum. 
Above 1\,GHz, the values fell to 2\,dB squeezing and 3\,dB anti-squeezing. The values and spectral behavior of the data from system\,2 were different. In the lower frequency range, we measured a noise suppression of about 6.5\,dB and up to 19\,dB anti-squeezing, which testifies to a significantly higher effective nonlinearity of the built-in crystal. Above 1\,GHz we measured 3.5\,dB squeezing and almost 5\,dB anti-squeezing. Our squeeze values were limited by optical loss, i.e.~by the imperfect quantum efficiency of our optical and photo-electrical system. 
Our anti-squeeze values were limited by the finite pump powers and for values at frequencies above 100\,MHz by the finite linewidths of the squeezing resonators.

\begin{figure}[h]
	\centering
	\begin{subfigure}{1\columnwidth} 
		{\includegraphics[width=\columnwidth]{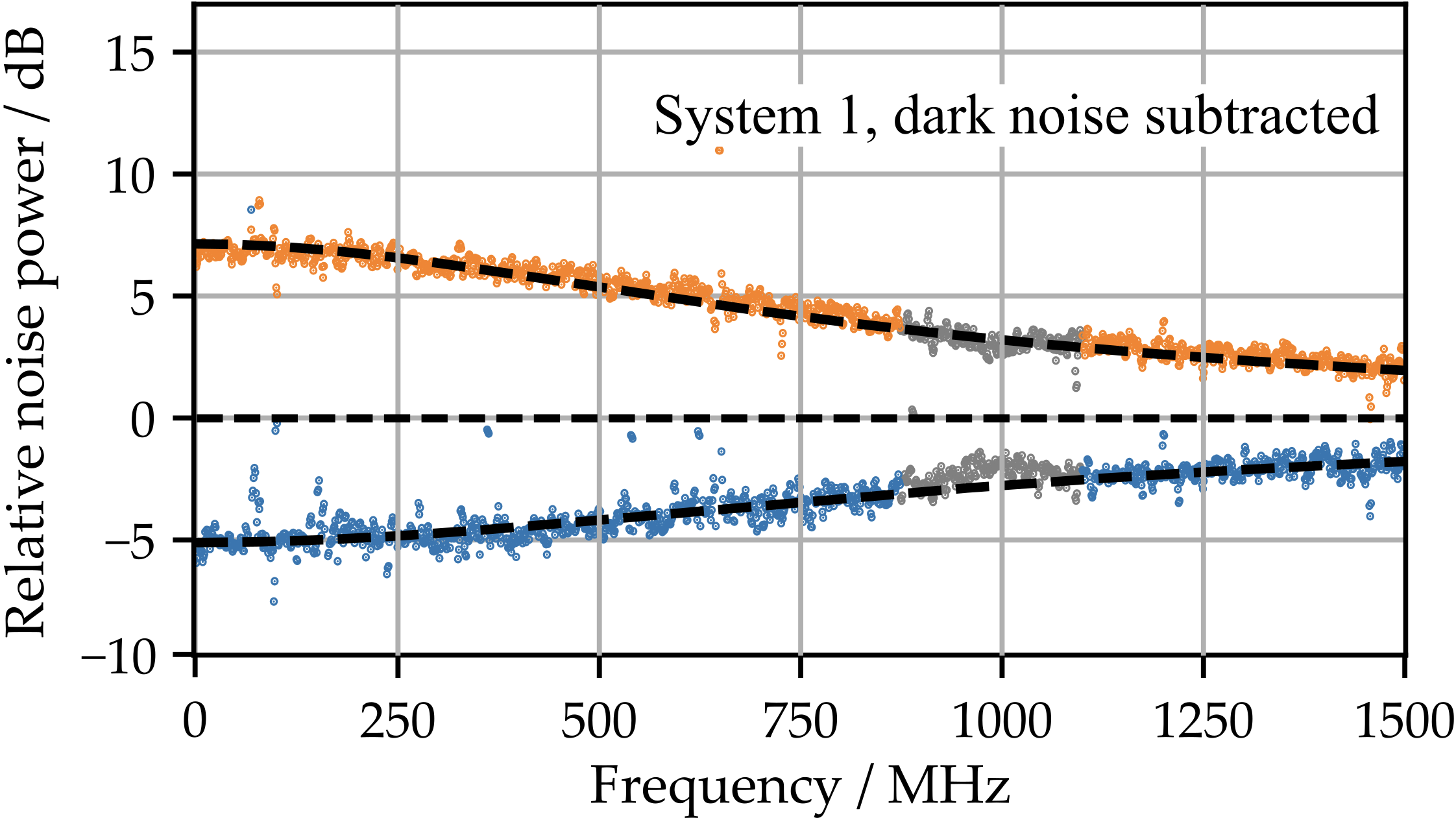}}
	\end{subfigure}
	\\[3mm]
	\begin{subfigure}{1\columnwidth} 
		{\includegraphics[width=\columnwidth]{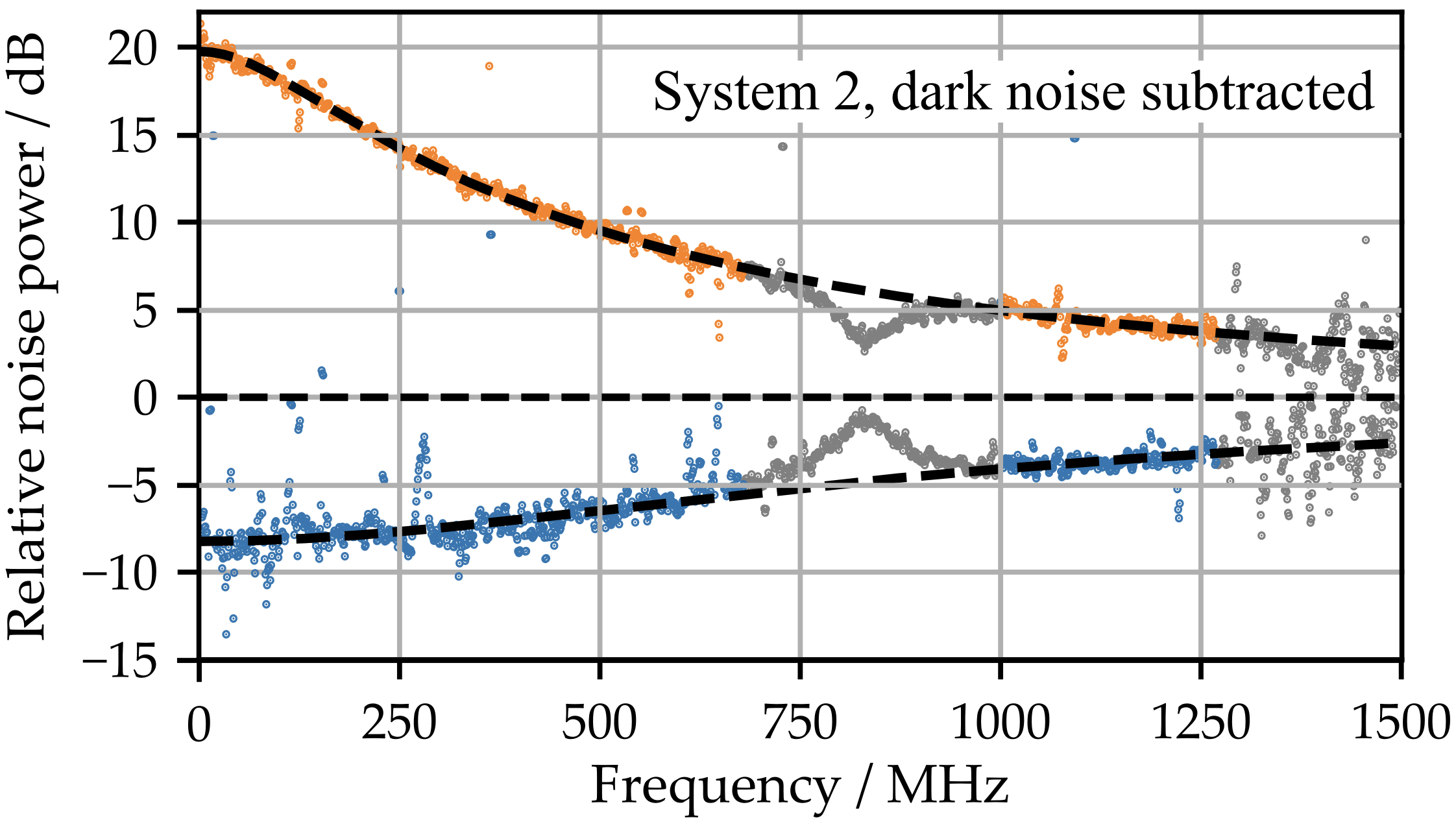}}
	\end{subfigure}  \vspace{-5mm}
	\caption{Shown is the same data as in Fig.\,\ref{fig:3}, but corrected for dark noise. The data points of all traces were converted to the linear scale and the dark noise was subtracted from the results, which were then returned to the dB scale. The dashed lines are analytical curves with fitted values for optical loss and squeezing resonator linewidths. The gray data were excluded from the fit. We assume that at these frequencies the phase delay of the detection electronics was unbalanced. After subtracting the dark noise, the squeeze factor in system\,2 increased from {\it measured} 6.5\,dB to {\it loss-limited} 8.5\,dB at around 30 MHz and from 3.5 to 4\,dB at around 1.1\,GHz respectively.}
	\label{fig:4}
	\vspace{2mm}
\end{figure}

Figure\;\ref{fig:4} presents the same measurement data as in Fig.\,\ref{fig:3}, but corrected for dark noise. The results are shot-noise-normalized traces that --- for the high parametric gain used --- are limited by the total optical efficiency of the setup including the measurement detection efficiency. This claim is supported for our data because of the excellent match with our theory with fitted values of resonator full-width-half-maximum linewidth $\gamma$ and total quantum efficiency $\eta$. The fitted values for system\,1 are $\,\gamma_1 = (2.05 \pm 0.06)\,$GHz and $\eta_1 = (82.8 \pm 1.6)$\%, and for system\,2 $\,\gamma_2 = (1.750 \pm 0.013)\,$GHz and $\eta_2 = (85.8 \pm 1.1)$\% \cite{Vahlbruch2016}. Grey data points were excluded from the fitting. The high quality of the fits also indicate that our BHDs were not saturating at the local oscillator powers of 8 and 6\,mW when {\it anti}-squeezed noise was measured. It should be noted that the linear scaling of {\it shot noise} with LO power shown in Fig.\,\ref{fig:2} does not necessarily guarantee the absence of saturation when anti-squeezed noise is measured with the same LO powers.

% loss
The efficiency values $\eta_1$ and $\eta_2$ must be compatible with an allocation to individual loss contributions. 
The supplier of the photodiodes (Laser Components) specified design values for the quantum efficiency of the photodiodes of approximately 99\%. Due to the small diameter of 100\,\textmu m, some light may be lost, so we assume $\eta_{\rm pd} \approx $\,98\% here. The mode matching efficiency with the local oscillator beams of the balanced homodyne detectors were $\eta_{\rm mm} \approx $\,97\%. During propagation, the squeezed beam suffered from scattering at lens surfaces, imperfect anti-reflection coatings and potential dust particles. Due to the compact setup we were not able to quantify the propagation efficiency precisely. For our consideration here, a rough estimation to $\eta_{\rm pg} \approx $\,97\% should be sufficient. The escape efficiency for a not damaged crystal can be estimated to  $\eta_{\rm ec} \approx $\,99\% \cite{Vahlbruch2016}. Our total efficiency values $\eta_1 \approx 83$\% and $\eta_2 \approx 86$\% should thus both correspond to the product of these values of about 91\%. The discrepancies are most likely due to crystal damage, in particular grey tracking caused by the intense pump field at 775\,nm. Grey tracking by such a long wavelength is rather unusual, however, our pump intensities are rather high, reaching 0.1\,kW  % 0.6W x 180
for a beam waist of approximately 33\,\textmu m. Note that the location of pump field (as well as the down-converted squeezed field) is fixed, since our squeezing resonators were monolithic (bi-convex) resonators.

We observed strongly different parametric gain versus pump power behaviors of SQZ\,1 and SQZ\,2, which was independent from potentially slightly different damages. Two effects likely contributed. As we had to use the crystal temperatures to keep the cavities on resonance, we could not adjust to optimal phase matchings independently. For squeeze laser 2 this lead to an operating temperature of \SI{29}{\degree C}, not too far off to the design phase matching temperature of \SI{40}{\degree C}, while we had to operate squeeze laser 1 at \SI{62}{\degree C} to achieve double resonance. The second effect contributing to the reduced non-linearity is the randomly positioned cut of the periodically poled crystal along the crystal axis. If there is a full poling period at one of the crystal ends, the non-linearity essentially cancels out to zero with each resonator revolution. Due to the monolithic structure of our resonators, we had no way of compensating for this with an adapted adjustment. 

The difference between the electronic noise of the homodyne detectors and the shot noise reference was 15\,dB and 10\,dB at 30\,MHz. Above 1\,GHz the detectors still have a dark noise clearance of 8\,dB and 9\,dB. A striking feature of our BHDs were the apparently reduced quantum efficiencies at 1\,GHz in SQZ\,1 and at 0.85\,GHz in SQZ\,2, see grey dots in Fig.\;\ref{fig:4}. An explanation would be that in both BHDs the two photodiodes together with the amplifier electronics generated different phases of the photovoltages due to resonances, so that neither the squeezed nor the anti-squeezed noise could be measured when subtracting the photovoltages, see also Fig.\;\ref{fig:1}. 

\section{Conclusion}
Our setup and measurements present the state-of-the-art production of GHz bandwidth squeezed states and their direct measurement in the nonclassical regime. By ``direct measurement'' we mean a measurement with a balanced homodyne detector (here of GHz bandwidth), without upstream optical parametric amplification into the semi-classical regime \cite{Shaked2018}. Our direct measurement scheme thus corresponds to the conventional one, in contrast to recent measurements with even much higher bandwidth \cite{Kashiwazaki2021}.  
We directly measure up to 6.5\,dB squeezing at 30\,MHz and still 3.5\,dB at 1.1\,GHz without dark noise subtraction. Both squeezing resonators were operated at the same time and with the same coherent pump laser, which opens the possibility to produce two-mode squeezed, EPR-entangled states with GHz bandwidth \cite{Furusawa1998, Eberle2011}, e.g.~for high speed one-sided device-independent QKD \cite{Gehring2015}.
Our squeezed light would also be suitable for improving laser interferometers for measuring gravitational waves in the speculative MHz to GHz frequency range \cite{Schnabel2024}.\\

\begin{backmatter}
\bmsection{Funding} The project is co-financed by ERDF of the European Union and by 'Fonds of the Hamburg Ministry of Science, Research, Equalities and Districts (BWFGB)'. SV was financially supported by the DFG unders Germany's Excellence Strategy EXC 2121 "Quantum Universe" -- 390833306. 

\bmsection{Disclosures} The authors declare no conflicts of interest.

\bmsection{Data availability} Data underlying the results presented in this paper are not publicly available at this time but may be obtained from the authors upon reasonable request.

\bigskip
\noindent

\end{backmatter}

\end{document}